\documentclass[11pt,preprint]{aastex}

\shorttitle{K Giant Diameters}
\shortauthors{Baines et al.}

\usepackage{amsmath}
\begin{document}


\title{Spectroscopic and Interferometric Measurements of Nine K Giant Stars}

\author{Ellyn K. Baines}
\affil{Remote Sensing Division, Naval Research Laboratory, 4555 Overlook Avenue SW, \\ Washington, DC 20375}
\email{ellyn.baines@nrl.navy.mil}

\author{Michaela P. D\"ollinger}
\affil{Max-Planck-Institut f\"ur Astronomie, K\"onigstuhl 17, D-69117 Heidelberg}

\author{Eike W. Guenther, Artie P. Hatzes}
\affil{Th\"uringer Landessternwarte Tautenburg, Sternwarte 5, D-07778 Tautenburg, Germany}

\author{Marie Hrudkovu}
\affil{Isaac Newton Group of Telescopes, Apartado de Correos 321, 387 00 Santa Cruz de la Palma, Canary Islands, Spain}

\author{Gerard T. van Belle}
\affil{Lowell Observatory, Flagstaff, AZ 86001} 

\begin{abstract}

We present spectroscopic and interferometric measurements for a sample of nine K giant stars. These targets are of particular interest because they are slated for stellar oscillation observations. Our improved parameters will directly translate into reduced errors in the final masses for these stars when interferometric radii and asteroseismic densities are combined. Here we determine each star's limb-darkened angular diameter, physical radius, luminosity, bolometric flux, effective temperature, surface gravity, metallicity, and mass. When we compare our interferometric and spectroscopic results, we find no systematic offsets in the diameters and the values generally agree within the errors. Our interferometric temperatures for seven of the nine stars are hotter than those determined from spectroscopy with an average difference of about 380 K.

\end{abstract}



\section{Introduction}

Giant stars are excellent candidates for both interferometric and asteroseismic observations. Interferometers have been used for many years to measure the angular diameters of giants, from the Mark III Interferometer \citep[e.g.,][]{2003AJ....126.2502M} to the Palomar Testbed Interferometer \citep[e.g.,][]{1999AJ....117..521V} to the Navy Precision Optical Interferometer \citep[e.g.,][]{1999AJ....118.3032N}. More recently, a sample of 25 K giant stars was measured by our team using the Center for High Angular Resolution Astronomy (CHARA) Array \citep{2010ApJ...710.1365B}. 

The other technique under consideration for this sample is asteroseismology, the study of stellar oscillations. It is a unique tool to infer the structure of stellar interiors with very little model dependence \citep[see, e.g.,][]{1994ARAandA..32...37B, 2004SoPh..220..137C}. Photometric space missions focusing on asteroseismology, i.e., \emph{MOST} \citep[\emph{Microvariability and Oscillations of STars,}][]{2003PASP..115.1023W}, \emph{CoRoT} \citep[\emph{Convection, Rotation, and planetary Transits,}][]{2006ESASP.624E..34B, 2009AandA...506..411A}, and \emph{Kepler} \citep{2013ApJ...765L..41S, 2010Sci...327..977B, 2010ApJ...713L..79K}, have dramatically increased both the number of stars with oscillation measurements as well as the quality of the data. These are critical measurements because the frequencies observed are dependent on the sound speed inside the star, which in turn depends on interior properties such as density, temperature, and gas motion \citep{2010AandA...509A..73C}. The stellar parameters resulting from these observations are key for testing stellar interior and evolutionary models \citep[see, e.g.,][]{2011Sci...332..213C}. 

Most giant stars, if not all of them, display measurable oscillations \citep[e.g.,][]{2013ApJ...765L..41S, 2006AandA...448..689D, 2002AandA...394L...5F, 1994ApJ...422..366H}, which makes them an ideal class of objects for deriving fundamental stellar parameters such as mass, radius, and temperature. They are bright, abundant, large enough to measure easily with interferometry, and exhibit radial velocity amplitudes from a few to several tens of m s$^{-1}$. The observed oscillation frequencies put constraints on the star's internal structure \citep{2006ApJ...647..558B}, namely the mean density of the star, while interferometry measures the star's size. The combination leads to the masses for these single stars. 

The defining characteristic of a star is its mass but for giant stars, determining this quantity is indirect and heavily model dependent. Often spectroscopic observations are used to measure a star's surface gravity (log~$g$), effective temperature ($T_{\rm eff}$), and iron abundance ([Fe/H]). The radius and mass are then determined by fitting evolutionary tracks to the star's position on the H-R diagram. This is an tricky process because the evolutionary tracks of stars with a large range of masses converge on the H-R diagram in the same region, and different evolutionary track models produce different masses for a given set of inputs. Without good calibrating objects, no set of tracks can be proven to be the best. Once we can test them by comparing theoretically determined mass and radius to measured values, we can have faith in applying the tracks to stars for which direct measurements are not possible. 

Several of the stars in our sample are ear-marked for asteroseismic studies using precise stellar radial velocity (PRV) measurements. It is difficult to obtain sufficient data in order to detect all pulsation modes using ground-based facilities. This requires a large amount of observing time often using multi-site campaigns. However, it is still possible to derive the stellar mass using a modest amount of ground-based data even taken at one site {\it if} one knows the stellar radius. This was done with some success for $\beta$ Gem \citep{2012AandA...543A..98H} and $\iota$ Dra \citep{2008AandA...491..531Z, 2011ApJ...743..130B}. PRV measurements will be made using the Thuringia State Observatory's 2 m telescope and McDonald Observatory's 2.7 m telescope, and results will be presented in a forthcoming paper. In the near future, network telescopes such as the Stellar Oscillations Network Group \citep[SONG,][]{2013POBeo..92...39G} should be able to investigate better the pulsations in these stars using PRVs. It is important, however, to first obtain stellar radii measurements, which is the goal of this paper.

The measured angular diameters, when combined with other measurements from the literature, ultimately lead to radii ($R$) and $T_{\rm eff}$ for the giant stars. These are important properties that characterize the star as well as the environment in which any possible exoplanets reside. Section 2 discusses the spectroscopic measurements of $T_{\rm eff}$, log $g$, and [Fe/H], section 3 describes the interferometric observations and calibrator star selection, section 4 outlines how we measure the angular diameter and calculate the $R$, luminosity, and $T_{\rm eff}$ for our sample, section 5 explores the physical implications of our measurements and plans for oscillation measurements, and section 6 summarizes our findings. 


\section{Spectroscopic Observations}

The sample of K giant stars presented here was obtained from the larger planet search survey of \citet{2007AandA...472..649D}. The stars chosen are bright ($V < 6.5$) K giants that show significant short-term variability indicative of stellar pulsations, which makes them perfect candidates for future asteroseismic measurements. 

The spectroscopic observations were obtained using the Coud\'e Echelle spectrograph of the 2-m Alfred Jensch telescope of the Th\"uringer Landessternwarte Tautenburg. The spectrograph has a resolving power of $\Delta\lambda/\lambda =67000$, and the wavelength range used was 4700 to 7400 \AA. Standard IRAF routines were used for subtracting the bias offset, flat-fielding, subtracting the scattered light, extracting the spectra, and for the wavelength calibration.\footnote{IRAF is distributed by the National Optical Astronomy Observatories, which are operated by the Association of Universities for Research in Astronomy, Inc., under cooperative agreement with the National Science Foundation.}

In order to determine $T_{\rm eff}$, log $g$, and [Fe/H] from the spectra, a grid of model atmospheres from \citet{1975AandA....42..407G} was used, which assumed a plane parallel atmosphere in local thermodynamic equilibrium. We used 144 unblended Fe\,I and 8 Fe\,II lines in the wavelength range 5806 and 6858 \AA\, using the line list of \citet{2004AandA...426..651P}. [Fe/H] was determined by assuming that Fe\,I lines of different equivalent widths have to give the same relative abundance of iron. For $T_{\rm eff}$, an excitation equilibrium of Fe\,I and Fe\,II for lines of different excitation potentials was used, and log $g$ was determined from the ionization balance of Fe\,I to Fe\,II lines \citep{doellinger08}. The radii were obtained by using Girardi evolutionary tracks \citep{2002AandA...391..195G} in their web-based form\footnote{http://stev.oapd.inaf.it/cgi-bin/param}. These tracks require a stellar magnitude, distance, $T_{\rm eff}$, and [Fe/H], and the output is the radius, mass, and age of the star. The resulting [Fe/H], $T_{\rm eff}$, log~$g$, and $R$ values are listed in Table~\ref{spec}.


\section{Interferometric Observations}

Interferometric observations were obtained using the CHARA Array, a six element optical/infrared 1-meter telescope array located on Mount Wilson, California \citep{2005ApJ...628..453T, 2005ApJ...628..439M}. We used the pupil-plane ``CHARA Classic'' beam combiner in the $K'$-band (2.133~$\mu$m), and the \emph{reduceir} pipeline written by T. ten Brummelaar\footnote{http://www.astro.gsu.edu/$\sim$theo/chara\_reduction/climb\_classic\_math.pdf} to reduce the data.

We interleaved data scans of the K giant stars with two to three calibrator stars for each target. We chose our calibrators to be stars that are significantly less resolved on the baselines used than the targets. This means that uncertainties in the calibrator's diameter do not affect the target's diameter calculation as much as if the calibrator had a substantial angular size. All scans were taken as close in time and space as possible, with preference given to calibrators within 7$^\circ$ of the targets, which was the case for all the target-calibrator pairings except for one with a separation of 12$^\circ$. We then converted instrumental target and calibrator measurements to calibrated data for the target stars.

To estimate the calibrator stars' angular diameters, we created spectral energy distribution (SED) fits to narrow- and wide-band photometric values published in \citet{1965ArA.....3..439L}, \citet{1981PDAO...15..439M}, \citet{1993AandAS..102...89O}, \citet{1990VilOB..85...50J}, \citet{1972VA.....14...13G}, \citet{1970AandAS....1..199H}, \citet{1991TrSht..63....1K}, \citet{1968tcpn.book.....E}, \citet{1966CoLPL...4...99J}, \citet{2003tmc..book.....C}, and \citet{1993cio..book.....G} as well as spectrophotometry from \citet{1983TrSht..53...50G}, \citet{1998yCat.3207....0G}, and \citet{1997yCat.3202....0K} obtained via the interface created by \citet{1997AandAS..124..349M}. The assigned uncertainties for the 2MASS infrared measurements are as reported in \citet{2003tmc..book.....C}, which in some cases are on the order of 0.3 mag, and an uncertainty of 0.05 mag was assigned to the optical measurements. Conversion from photometric magnitudes to fluxes incorporated zero-point uncertainties associated with the photometric systems as found in their reference literature \citep[e.g., see discussions in][]{1995PASP..107..945F, 2015PASP..127..102M} and are on order 2$\%$. 

The flux-calibrated stellar templates of \citet{1998PASP..110..863P} were chosen based on each star's spectral type and fit to the photometry. The templates were then adjusted to account for the overall flux level and reddening, and to estimate angular diameter using the $\chi^2$ minimization technique. The resulting SEDs gave us each star's bolometric flux ($F_{\rm BOL}$) and allowed for the calculation of extinction ($A_{\rm V}$) using the wavelength-dependent reddening relations of \citet{1989ApJ...345..245C}, assuming a `standard' R$_{\rm V}$=3.1 wavelength progression of reddening. The SED fits allowed us to check if there was any excess emission that might be due to an otherwise unknown low-mass companion or circumstellar disk. Any calibrator candidates displaying variable radial velocities, photometric variations, or any indication of binarity were discarded. Table \ref{obs_log} lists the K giant stars observed, the date and baseline used, and calibrator information.

We observed every target with multiple calibrator stars to check on the calibrators themselves. We used calibrator 1 as a check for calibrator 2 and vice versa, and used them individually as well as in conjunction to measure the angular diameter of the target star. These results were consistent, whether we used one or the other calibrator or both together, so there do not appear to be any systematics in the data arising from the calibrators themselves.


\section{Results}

\subsection{Angular Diameter Measurement}

We fit measured calibrated visibilities ($V$), the observed quantity of an interferometer, to both uniform disk (UD) and limb darkened (LD) angular diameters ($\theta$). For more details on this procedure, see \citet{1974MNRAS.167..475H, 1992ARAandA..30..457S, 2010ApJ...710.1365B}. The uncertainties on $V$ consist of several parts combined in quadrature: the formal error on the mean of the visibility measurement; the amount the calibrator's visibilities changes over the course of the observations; and the calibrator diameters and their associated uncertainties. These are are taken into account using the calibration process described in \citet{2005PASP..117.1263V}.

\begin{figure}[h]
\includegraphics[width=1.0\textwidth]{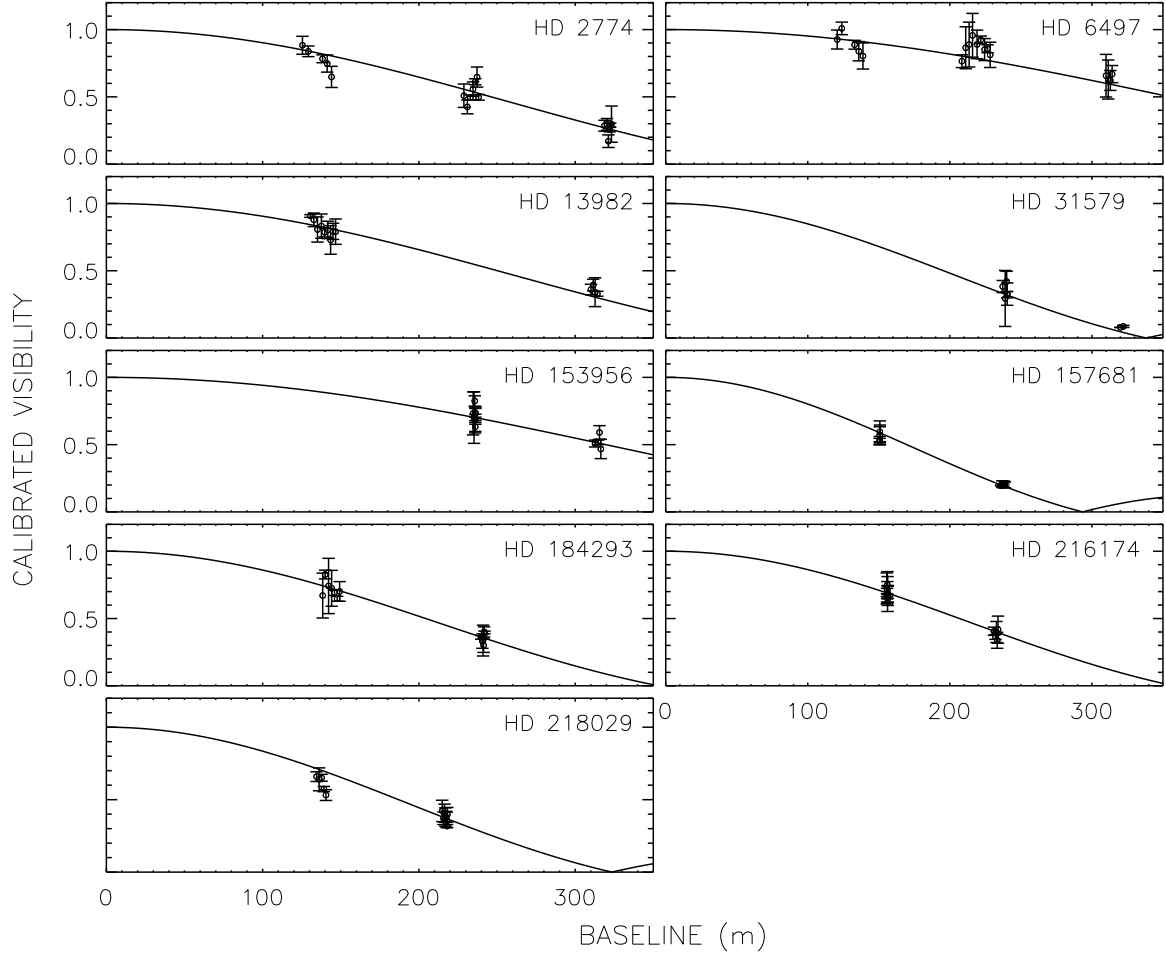}
\vspace{-1.5in}
\caption{$\theta_{\rm LD}$ fits for the nine K giant stars. The solid lines represent the visibility curve for the best fit $\theta_{\rm LD}$, the points are the calibrated visibilities, and the vertical lines are the measurement uncertainties.}
  \label{ldplot}
\end{figure}

The conversion between UD and LD diameters involves the LD coefficient ($\mu_{\lambda}$) from \citet{2011AandA...529A..75C}, which was obtained using $T_{\rm eff}$, log $g$, and [Fe/H] values from the spectroscopic observations with a microturbulent velocity of 2 km s$^{\rm -1}$. The average difference between the $\theta_{\rm UD}$ and $\theta_{\rm LD}$ are on the order of a few percent, and the final $\theta_{\rm LD}$ is little affected by the choice of $\mu_{\lambda}$: a 20$\%$ change in $\mu_{\lambda}$ results in at most a 1$\%$ change in $\theta_{\rm LD}$. All stars have errors in $\theta_{\rm LD}$ 1 to 3$\%$, except for HD 6497, which has an error of 6$\%$ and is the star with the smallest angular diameter. Table \ref{inf} lists $\theta_{\rm UD}$, $\mu_{\lambda}$, and $\theta_{\rm LD}$. Figure \ref{ldplot} shows the $\theta_{\rm LD}$ fits for all the stars. The calibrated visibilities are available in the online version of \emph{The Astronomical Journal}.

For each $\theta_{\rm LD}$ fit, the errors were derived via the reduced $\chi^2$ minimization method \citep{2003psa..book.....W,1992nrca.book.....P}: the diameter fit with the lowest $\chi^2$ was found and the corresponding diameter was the final $\theta_{\rm LD}$ for the star. The errors were calculated by finding the diameter at $\chi^2 \pm 1$ on either side of the minimum $\chi^2$ and determining the differences between the $\chi^2$ diameter and $\chi^2 \pm 1$ diameters. The reduced $\chi^2$ were between 2 and 4 for all the stars, and when $\chi^2$ was forced to equal one, the errors increased. We used the larger errors to be on the conservative side, and these are the errors listed in Table \ref{inf}.

\subsection{Stellar Radius, Luminosity, and Effective Temperature}

We combined our $\theta_{\rm LD}$ measurements with \emph{Hipparcos} parallaxes \citep{2007AandA...474..653V} to calculate the stars' $R$. In order to determine the luminosity ($L$) and $T_{\rm eff}$, we used the procedure described in section 3 to create SED fits. We combined our $F_{\rm BOL}$ values with the stars' distances ($d$) to estimate $L$ using $L = 4 \pi d^2 F_{\rm BOL}$. We also combined the $F_{\rm BOL}$ with $\theta_{\rm LD}$ to determine each star's $T_{\rm eff}$ by inverting the relation,
\begin{equation}
F_{\rm BOL} = {1 \over 4} \theta_{\rm LD}^2 \sigma T_{\rm eff}^4,
\end{equation}
where $\sigma$ is the Stefan-Boltzmann constant and $\theta_{\rm LD}$ is in radians. 

Considering that $\mu_\lambda$ is selected based on a given $T_{\rm eff}$, we checked to see if $\mu_\lambda$ and the resulting $\theta_{\rm LD}$ changed based on our new $T_{\rm eff}$. When selecting the updated $\mu_\lambda$ using our measured $T_{\rm eff}$, the largest difference in $\mu_\lambda$ was 0.04, which was the case for three stars, and was $\leq$0.02 for the remainder. The resulting $\theta_{\rm LD}$ values changed at most by 0.5$\%$, and all but one changed by 0.3$\%$ or less. This was well within the uncertainties on $\theta_{\rm LD}$, and re-calculating $T_{\rm eff}$ with the new $\theta_{\rm LD}$ made at most a 14 K difference. The $T_{\rm eff}$ values all converged after this one iteration, and these are the final values listed in Table \ref{inf}. Metallicity had a small effect on $\mu_\lambda$ and the final $\theta_{\rm LD}$: we varied the metallicity by $\pm$1.0 and recalculated the $\mu_\lambda$ and $\theta_{\rm LD}$. It made at most a 0.003 mas change in the final diameters, which is within the errors.

\section{Discussion}

\subsection{Comparing Spectroscopic and Interferometric Diameters}

\begin{figure}[h]
\includegraphics[width=1.0\textwidth]{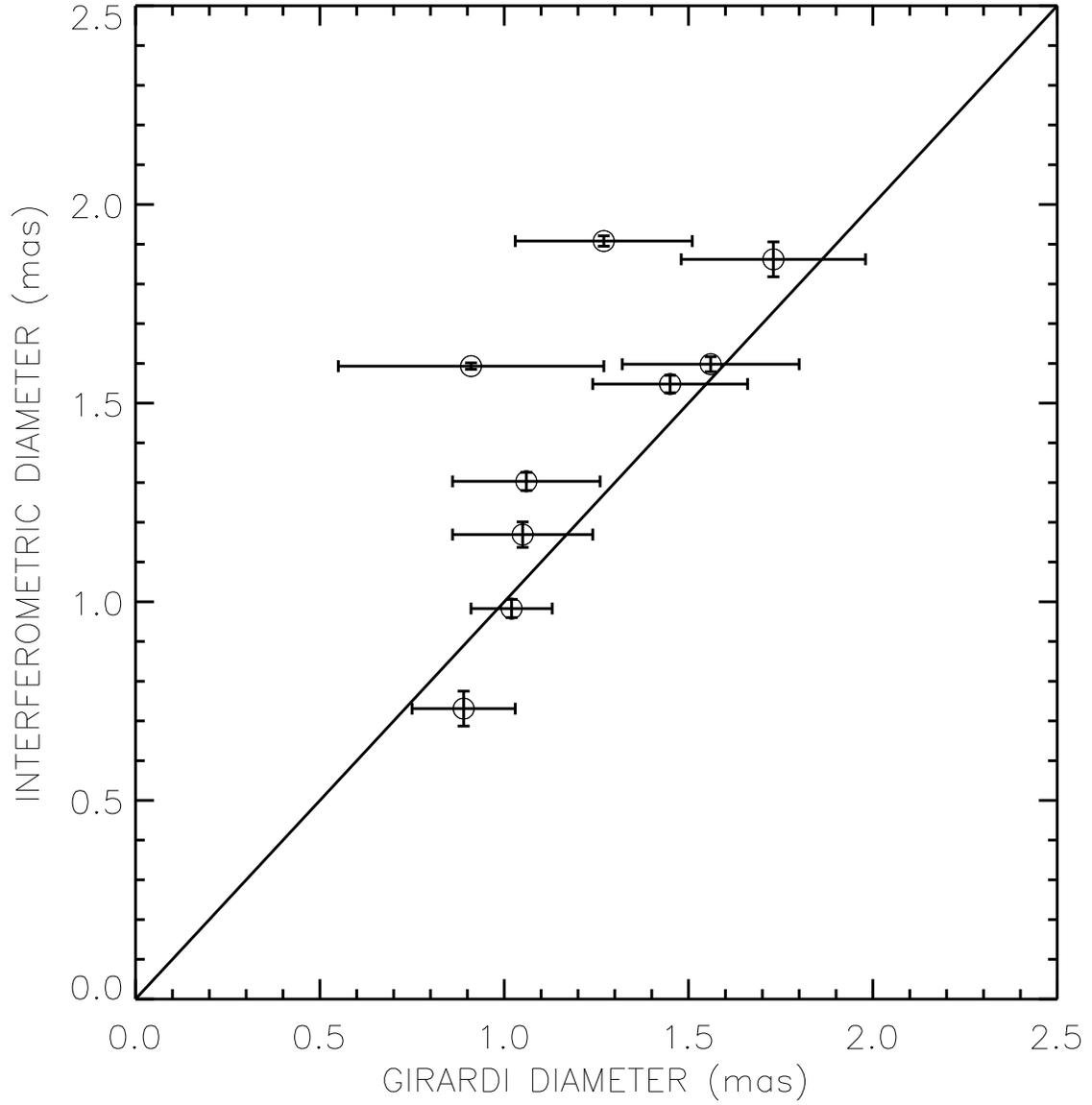}
\caption{A comparison of spectroscopically estimated versus interferometrically measured angular diameters. The solid line is the 1:1 ratio between the two quantities. The largest outliers are HD 31579 and HD 157681. See $\S$5 for a discussion on these stars.}
  \label{angdiam}
\end{figure}

We compared the angular diameters predicted using the Girardi tracks using spectroscopically determined $T_{\rm eff}$ and [Fe/H] against the interferometric measurements in Figure \ref{angdiam}. For the most part, the diameters agree within the errors and there is no clear bias. The error bars on the interferometric measurements are substantially smaller than those on the Girardi diameters, between 3$\times$ and 19$\times$ smaller: the errors for $\theta_{\rm interf}$ are on the order of 1 to 3$\%$ with just one at 6$\%$, while the errors for $\theta_{\rm Girardi}$ range from 11$\%$ to 18$\%$.

The largest outliers in Figure \ref{angdiam} are HD 31579 and HD 157681. The latter was observed as part of \citet{2010ApJ...710.1365B}, and its interferometric diameter of 1.664$\pm$0.010 mas was larger than the diameter predicted by spectroscopy (1.27$\pm$0.24 mas). Baines et al. concluded it was likely due to the calibrator star used (HD 158460) so we observed it again using two different calibrators. Our new diameter of 1.901$\pm$0.013 mas is even larger than the previous measurement. However, when the data are analyzed using each calibrator star separately, the resulting angular diameters are remarkably consistent with a mere 0.003 mas difference. When the calibrators are used to calibrate each other, no systematic offsets are present. We also used the relationship described in \citet{1999PASP..111.1515V} between the angular diameter and the ($V-K$) color to estimate HD 157681's diameter and obtained 2.05$^{+0.45}_{-0.82}$ mas, which agrees with our new interferometric measurement to within the errors.

As for HD 31579, the spectroscopically determined angular diameter (0.91$\pm$0.36 mas) is the outlier when considered against the those determined using the SED fit (1.60$\pm$0.12 mas), the ($V-K$) color (1.67$\pm$0.27 mas), and the interferometric measurement (1.593$\pm$0.008 mas). All the diameters are consistent and agree to within the errors except for the spectroscopic calculation.

\subsection{Comparing Spectroscopic and Interferometric Temperatures}

\begin{figure}[h]
\includegraphics[width=1.0\textwidth]{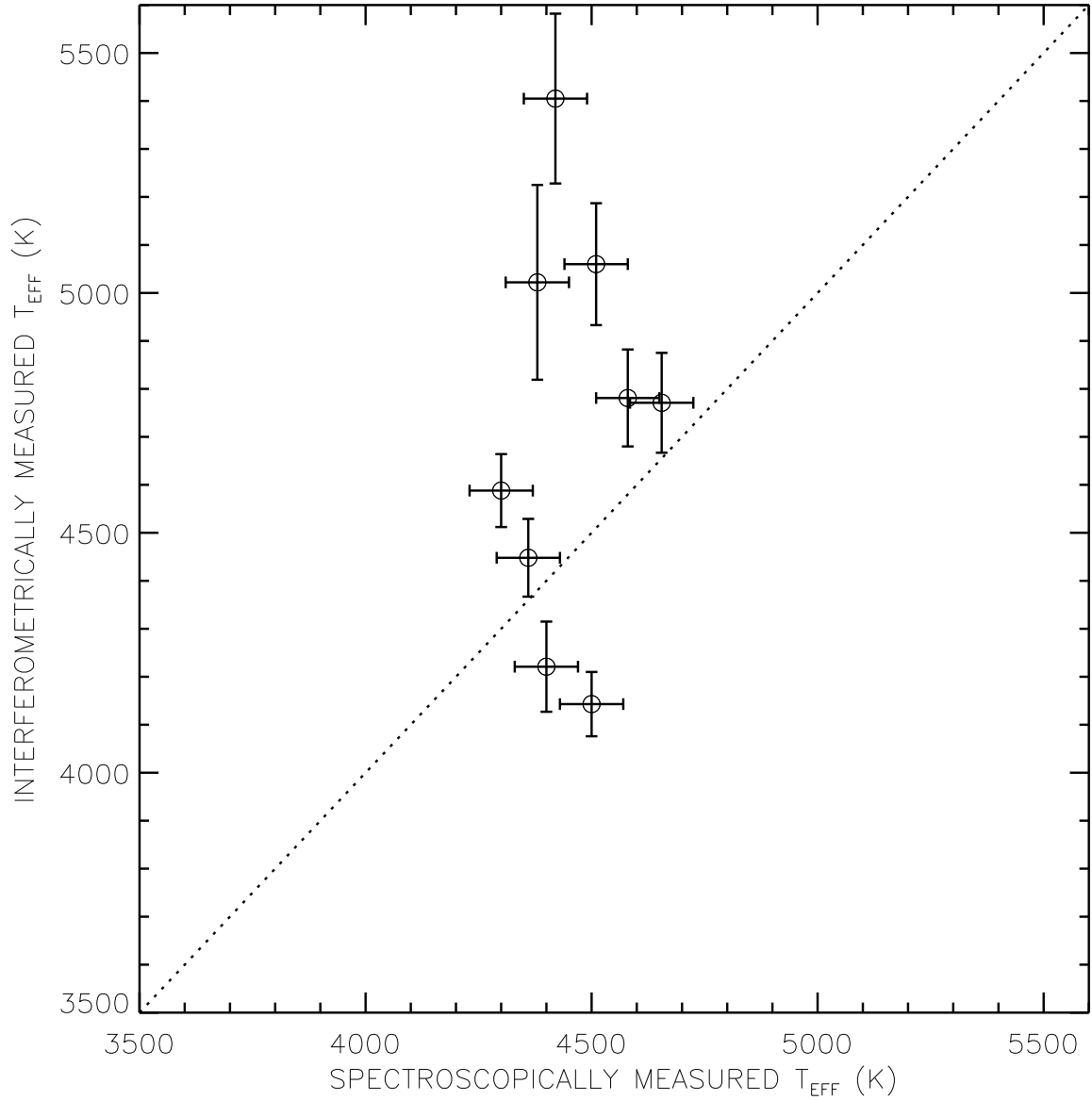}
\caption{A comparison of spectroscopically and interferometrically measured $T_{\rm eff}$. The dotted line is the 1:1 ratio, and the solid line is a linear fit to the data.}
  \label{tcompare}
\end{figure}

We plotted the spectroscopically determined $T_{\rm eff}$ versus our interferometric results in Figure \ref{tcompare}. There is some scatter off the 1:1 line with the spectroscopic values tending to be cooler than the interferometric ones by an average of $\sim$380 K. The discrepancy may be due to the atmospheric models of K giant stars in the near-ultraviolet lacking a source of thermal extinction, which could affect the $T_{\rm eff}$ measurements \citep{2009ApJ...691.1634S}. Another cause may lie in the methods used to determine $T_{\rm eff}$: interferometry measures the overall $T_{\rm eff}$ of the star while spectroscopic values rely on Fe\,I and Fe\,II lines and measure the $T_{\rm eff}$ in the thin layers of the atmosphere where those lines are formed. In dwarf stars, local thermodynamic equilibrium is a reasonable assumption and the $T_{\rm eff}$ determined using the iron lines is the same as the $T_{\rm eff}$ of the atmosphere overall. For giant stars, the atmosphere is more extended and the models may not be correct due to factors such as convection. Another consideration may be that the 1-D models do not include geometrical surface cooling and the 2-D models may not be as extended as real stars, so do not perfectly describe the atmospheres.

HD 157681 is again an object of interest when it comes to determining its $T_{\rm eff}$. In order to match $\theta_{\rm spec}$, $T_{\rm eff}$ would have to drop from 4400 K to 3844 K, which is much closer to the 3900 K predicted by the ($B-V$) color. This has the effect of moving the star from below the 1:1 line in Figure \ref{tcompare} to above it, which is consistent with the rest of the stars except for HD 31579. HD 157681 is the coolest giant in the sample, which is expected because it is a K5 star while the others are K0 to K3.

As an independent check on $T_{\rm eff}$, we used the equations from \citet{2010MNRAS.403.1592B} that relate ($B-V$) color, bolometric correction (BC$_V$), and $T_{\rm eff}$ for stars between 3300 and 5000 K. The results are listed in Table \ref{tcompare_table}. Color $T_{\rm eff}$ are even cooler than the spectroscopic $T_{\rm eff}$, except for HD 216174 where they are equal. On average the spectroscopic $T_{\rm eff}$ are hotter than the color $T_{\rm eff}$ by $\sim$320 K, while the interferometric $T_{\rm eff}$ are hotter on average by $\sim$580 K. We also did a search in the literature using the VizieR service and averaged all available $T_{\rm eff}$ values, and these are included in Table \ref{tcompare_table}. 

\begin{figure}[h]
\includegraphics[width=1.0\textwidth]{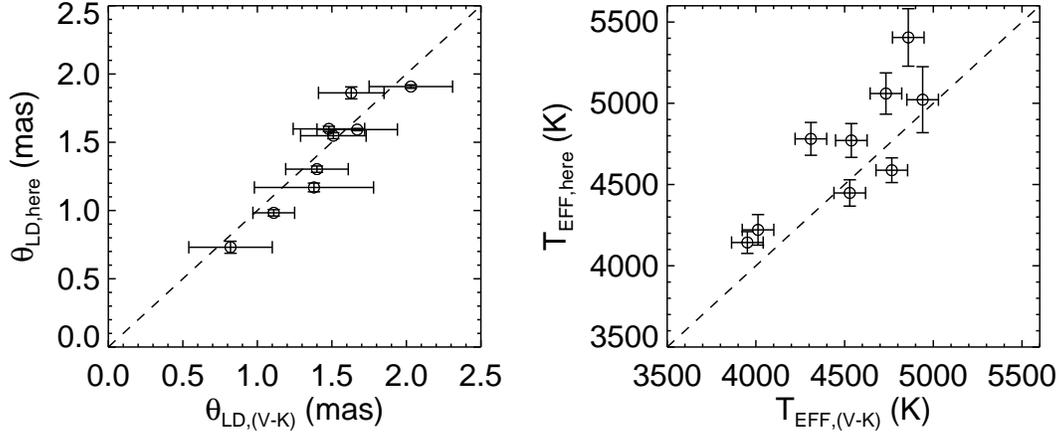}
\caption{A comparison of $\theta_{\rm LD}$ (left panel) and $T_{\rm eff}$ (right panel) using interferometric measurements and the procedure described in Section 5.1. The dashed line is the 1:1 ratio between the two quantities. The $T_{(V-K)}$ errors are 89 K, which is the standard deviation of the residuals as noted in \citet{2003AJ....126.2502M}.}
  \label{diams_teff}
\end{figure}

As a final check, we calculated both $\theta_{\rm LD}$ and $T_{\rm eff}$ using the relations between them and the surface brightness and ($V-K$) color, respectively, described in \citet{2003AJ....126.2502M}. Table \ref{tcompare_table} lists the resulting values, which are also plotted in Figure \ref{diams_teff}. The diameters show a scatter around the 1:1 ratio but are within the errors, and we see a similar offset in $T_{\rm eff}$, where our new measurements are hotter than those predicted using Mozurkewich et al.'s equations for seven of the nine stars. When we compare the temperatures determined spectroscopically, interferometrically, and using the ($V-K$) colors, four of the nine stars have $T_{\rm inf}$ that fall in between the $T_{\rm spec}$ and $T_{(V-K)}$. 

\subsection{Future Oscillation Studies}

The velocity amplitude of the K giant stars' p-mode oscillations range from a few to tens of m\,s$^{-1}$, depending on the evolutionary state of the star \citep{1995AandA...293...87K}. The mode periods range from several hours to days. These amplitudes and periods are measureable with 2-3 m class telescopes using precise stellar RV measurements, which typically reach a precision of $\sim$1 m\,s$^{-1}$.

We intend to use the Coud\'e echelle spectrograph of the 2-m Alfred Jensch Telescope of the Thuringia State Observatory to detect the stellar oscillations in those stars for which we have interferometrically measured $R$. An iodine absorption cell will be used to provide the wavelength calibration for the RV measurement. This instrument is able to achieve and RV precision of $\sim$2 m\,s$^{-1}$ on bright K giant stars \citep{2012AandA...543A..98H}.

Fundamental stellar parameters of K giant stars are important for exoplanet studies because of their masses, which can be 1.5--3 $M_\odot$. Main-sequence stars of this mass range are ill-suited for RV measurements due to a paucity of stellar lines and high stellar rotation rates. Thus K giants offer us a means to study planet formation around stars more massive than the Sun. 

\subsection{Stellar Masses}

Determining $M$ for these giant stars is key to understanding whether or not planet populations orbiting massive stars are different than planets found orbiting solar-type stars. Some scientists argue that more massive stars host more massive planets, and that A stars are at least five times more likely to host a giant planet than an M dwarf \citep{2010ApJ...709..396B, 2010PASP..122..149J, 2010PASP..122..701J, 2012AandA...544A...9V}. There are models that support this theory: e.g., \citet{2008ApJ...673..502K, 2013ApJ...778...78H}. However, \citet{2011ApJ...739L..49L, 2013ApJ...774L...2L} disagrees, claiming the masses determined for the exoplanet host stars are in error due to the convergence and crossing of evolutionary tracks from stellar models. This leads to degeneracies, and Lloyd believes the masses of the evolved stars are not as high as those claimed by previous studies.

Our ultimate contribution to this controversy will be the direct determination of $M$ for our sample of giant stars by combining our interferometric $R$ with the asteroseismic density measurements. We will then be able to determine if the models are indeed correct, and test if the idea that more massive stars host more massive planets is valid.


\section{Summary}

We measured the angular diameters of nine K giant stars that are the targets for future exoplanet searches and asteroseismology studies. We combined our measurements with information from the literature to calculate each star's $R$ and $T_{\rm eff}$, and used SED fits to determine $L$ and $F_{\rm BOL}$.

Our improved angular diameter precision translates directly to smaller errors when calculating the physical radii for these targets, which will in turn lead to reduced errors when determining the mass from stellar oscillation studies. Once those masses have been measured, we can compare them to results from evolutionary models to help distinguish between which isochrones best match our observations. Those models can then be applied to stars for which interferometric or asteroseismic measurements are not possible.

\acknowledgments

This work is based upon observations obtained with the Georgia State University Center for High Angular Resolution Astronomy Array at Mount Wilson Observatory. The CHARA Array is supported by the National Science Foundation under Grant No. AST-1211929. Institutional support has been provided from the GSU College of Arts and Sciences and the GSU Office of the Vice President for Research and Economic Development. APH, MP, and MD acknowledge DFG grants HA 3279/5-1 and HA 3279/9-1. We are also grateful to the user support group of the Alfred-Jensch telescope. This research has made use of the SIMBAD database, operated at CDS, Strasbourg, France. This publication makes use of data products from the Two Micron All Sky Survey, which is a joint project of the University of Massachusetts and the Infrared Processing and Analysis Center/California Institute of Technology, funded by the National Aeronautics and Space Administration and the National Science Foundation.


\begin{deluxetable}{llllrcccrr}
\tablewidth{0pc}
\tabletypesize{\scriptsize}
\tablecaption{Observed and Spectroscopic Properties. \label{spec}}

\tablehead{\colhead{Target} & \colhead{$V$} & \colhead{$K$} & \colhead{Spectral} & \colhead{$\pi$} & \colhead{$T_{\rm eff}$} & \colhead{log~$g$}  & \colhead{[Fe/H]} & \colhead{$\theta_{\rm spec}$} & \colhead{$R_{\rm spec}$} \\ 
           \colhead{HD}     & \colhead{mag} & \colhead{mag} & \colhead{Type} & \colhead{(mas)} & \colhead{$\pm$70 K}   & \colhead{$\pm$0.2} & \colhead{$\pm$0.5 dex}    & \colhead{(mas)}        & \colhead{($R_\odot$)}       }
\startdata
2774   & 5.59 & 2.80$\pm$0.09$^a$ & K2 III &  8.56$\pm$0.41  & 4655 & 2.7 & -0.08 & 1.06$\pm$0.20 & 13.73$\pm$1.48 \\
6497   & 6.42 & 3.88$\pm$0.34$^b$ & K2 III & 10.09$\pm$0.52  & 4420 & 2.4 & -0.08 & 0.89$\pm$0.14 &  9.30$\pm$0.82 \\
13982  & 5.75 & 2.88$\pm$0.32$^b$ & K3 III &  7.94$\pm$0.44  & 4580 & 2.3 & -0.07 & 1.05$\pm$0.19 & 13.09$\pm$1.37 \\
31579  & 6.08 & 2.63$\pm$0.09$^a$ & K3 III &  5.67$\pm$0.62  & 4500 & 2.8 & +0.06 & 0.91$\pm$0.36 & 23.01$\pm$4.47 \\
153956 & 6.03 & 3.28$\pm$0.34$^b$ & K1 III & 10.74$\pm$0.55  & 4510 & 2.3 & -0.08 & 1.02$\pm$0.11 &  9.96$\pm$0.65 \\
157681 & 5.67 & 2.19$\pm$0.05$^a$ & K5 III &  5.23$\pm$0.27  & 4400 & 1.6 & -0.23 & 1.27$\pm$0.24 & 24.66$\pm$2.47 \\
184293 & 5.53 & 2.59$\pm$0.06$^a$ & K1 III &  7.06$\pm$0.22  & 4380 & 1.9 & -0.26 & 1.45$\pm$0.21 & 22.28$\pm$1.69 \\
216174 & 5.38 & 2.64$\pm$0.06$^a$ & K1 III &  8.21$\pm$0.25  & 4300 & 1.2 & -0.55 & 1.56$\pm$0.24 & 19.14$\pm$1.62 \\
218029 & 5.25 & 2.48$\pm$0.05$^a$ & K3 III &  7.89$\pm$0.22  & 4360 & 2.0 & +0.07 & 1.73$\pm$0.25 & 21.87$\pm$1.78 \\
\enddata
\tablecomments{$^a$\emph{Two-Micron Sky Survey} \citep{1969tmss.book.....N}; $^b$\emph{2MASS All-Sky Catalog of Point Sources} \citep{2003tmc..book.....C}; $V$ magnitudes are from \citet{Mermilliod};  parallaxes ($\pi$) are from \citet{2007AandA...474..653V}; spectral types, $T_{\rm eff}$, log~$g$, [Fe/H], $\theta_{\rm spec}$, and $R_{\rm spec}$ are from \citet{doellinger08}.}
\end{deluxetable}

\clearpage


\begin{deluxetable}{ccccccccccc}
\tablewidth{0pc}
\tabletypesize{\scriptsize}
\tablecaption{Observing Log and Calibrator Star Information.\label{obs_log}}
\tablehead{\multicolumn{5}{c}{Observing Log} & \multicolumn{1}{c}{ } & \multicolumn{5}{c}{Calibrator Information} \\
\cline{1-5} \cline{7-11} \\
 \colhead{Target} & \colhead{Calibrator} & \colhead{Date} & \colhead{Baselines}      & \colhead{$\#$} & \colhead{ } & \colhead{$T_{\rm eff}$} & \colhead{log~$g$}       & \colhead{ }   & \colhead{$A_{\rm V}$} & \colhead{$\theta_{\rm est}$} \\
 \colhead{HD}     & \colhead{HD}         & \colhead{(UT)} & \colhead{Used$^\dagger$} & \colhead{Obs}  & \colhead{ } & \colhead{(K)}           & \colhead{(cm s$^{-2}$)} & \colhead{Ref} & \colhead{(mag)}       & \colhead{(mas)}            \\ }
\startdata
2774   &   4222 & 2010/07/29 & S2-E2 &  7 & &  9000 & 4.21 & 1 & 0.16$\pm$0.02 & 0.32$\pm$0.02 \\
       &        & 2010/08/01 & W2-E2 &  3 & &       &      &   &               &                 \\
       &        & 2013/09/02 & S1-E1 &  1 & &       &      &   &               &                 \\
       &        & 2013/09/04 & S1-E1 &  4 & &       &      &   &               &                 \\
       &   6961 & 2010/07/29 & S2-E2 &  7 & &  7762 & 3.80 & 2 & 0.02$\pm$0.02 & 0.55$\pm$0.04 \\
       &        & 2010/08/01 & W2-E2 &  4 & &       &      &   &               &                 \\
       &        & 2013/09/02 & S1-E1 &  1 & &       &      &   &               &                 \\
       &        & 2013/09/04 & S1-E1 &  6 & &       &      &   &               &                 \\
6497   &   4222 & 2010/07/29 & S2-E2 & 10 & &       &      &   &               &                 \\
       &        & 2010/08/01 & W2-E2 &  5 & &       &      &   &               &                 \\
       &        & 2013/09/02 & S1-E1 &  5 & &       &      &   &               &                 \\
       &   6961 & 2010/07/29 & S2-E2 & 10 & &       &      &   &               &                 \\
       &        & 2010/08/01 & W2-E2 &  5 & &       &      &   &               &                 \\
       &        & 2013/09/02 & S1-E1 &  5 & &       &      &   &               &                 \\
13982  &  11151 & 2010/08/01 & W2-E2 &  9 & &  6761 & 4.12 & 2 & 0.02$\pm$0.02 & 0.46$\pm$0.03 \\
       &  12303 & 2010/08/01 & W2-E2 &  9 & & 11100 &  3.4 & 3 & 0.33$\pm$0.01 & 0.27$\pm$0.02 \\
       &        & 2013/09/04 & S1-E1 &  4 & &       &      &   &               &                 \\
       &  20365 & 2013/09/04 & S1-E1 &  3 & & 19000 & 3.94 & 1 & 0.58$\pm$0.03 & 0.19$\pm$0.01 \\
31579  &  29526 & 2016/02/10 & S2-E2 &  5 & &  9550 & 4.12 & 2 & 0.17$\pm$0.02 & 0.23$\pm$0.02 \\
       &  33167 & 2016/02/10 & S2-E2 &  5 & &  6607 & 3.96 & 2 & 0.09$\pm$0.02 & 0.52$\pm$0.04 \\
       &  38091 & 2016/02/11 & S1-E1 &  2 & &  8128 & 4.26 & 2 & 0.07$\pm$0.02 & 0.38$\pm$0.02 \\
       &  46590 & 2016/02/11 & S1-E1 &  2 & &  9550 & 4.14 & 2 & 0.06$\pm$0.02 & 0.24$\pm$0.02 \\
153956 & 151044 & 2010/07/29 & S2-E2 &  4 & &  6166 & 4.38 & 2 & 0.04$\pm$0.02 & 0.40$\pm$0.03 \\
       & 158460 & 2010/07/29 & S2-E2 & 10 & &  9395 & 4.19 & 1 & 0.14$\pm$0.02 & 0.27$\pm$0.02 \\
       &        & 2013/09/02 & S1-E1 &  4 & &       &      &   &               &                 \\
157681 & 158414 & 2010/07/30 & S2-E2 &  5 & &  8000 & 4.24 & 1 & 0.52$\pm$0.02 & 0.40$\pm$0.03 \\
       &        & 2010/07/31 & W2-E2 &  5 & &       &      &   &               &                 \\
       & 161693 & 2010/07/30 & S2-E2 &  5 & &  9000 & 4.19 & 1 & 0.15$\pm$0.02 & 0.26$\pm$0.01 \\
       &        & 2010/07/31 & W2-E2 &  5 & &       &      &   &               &                 \\
184293 & 184006 & 2010/08/01 & W2-E2 &  5 & &  8180 & 4.29 & 1 & 0.13$\pm$0.02 & 0.71$\pm$0.05 \\
       & 184960 & 2010/07/30 & S2-E2 &  8 & &  6457 & 4.33 & 2 & 0.00$\pm$0.01 & 0.56$\pm$0.04 \\
       &        & 2010/08/01 & W2-E2 &  7 & &       &      &   &               &                 \\
216174 & 212454 & 2010/07/30 & S2-E2 &  3 & & 15750 & 4.20 &   & 0.04$\pm$0.02  & 0.11$\pm$0.01 \\
       &        & 2010/07/31 & W2-E2 &  9 & &       &      &   &               &                \\
       & 218470 & 2010/07/30 & S2-E2 &  7 & &  6761 & 4.21 & 2 & 0.07$\pm$0.02 & 0.51$\pm$0.04 \\
       &        & 2010/07/31 & W2-E2 &  9 & &       &      &   &               &                 \\
218029 & 219485 & 2010/07/30 & S2-E2 &  8 & &  9790 & 4.14 & 1 & 0.00$\pm$0.02 & 0.23$\pm$0.02 \\
       &        & 2010/08/01 & W2-E2 &  5 & &       &      &   &               &                 \\
       & 223274 & 2010/07/30 & S2-E2 &  8 & &  9120 & 3.80 & 2 & 0.05$\pm$0.02 & 0.34$\pm$0.02 \\
       &        & 2010/08/01 & W2-E2 &  5 & &       &      &   &               &                \\
\enddata
\tablecomments{$^\dagger$The maximum baseline lengths are W2-E2 156 m, S2-E2 248 m, and S1-E1 331 m. \\
References: (1) \citet{cox00}, based on spectral type as listed in the \emph{SIMBAD Astronomical Database}; (2) \citet{1999AandA...352..555A}; (3) \citet{2010yCat.2300....0L}; (4) \citet{2005ApJS..159..141V}; (5) \citet{1997AandAS..124..299C}. The estimated angular diameter $\theta_{\rm est}$ and $A_{\rm V}$ was determined using the fitting procedure described in $\S$3.}
\end{deluxetable}

\clearpage


\begin{deluxetable}{lcccccrcclcc}
\rotate
\tablewidth{0pc}
\tabletypesize{\scriptsize}
\tablecaption{Stellar Parameters. \label{inf}}

\tablehead{\colhead{Target} &  \colhead{$\theta_{\rm UD,inf}$} & \colhead{$\mu_{\rm \lambda}$} & \colhead{$\mu_{\rm \lambda}$} & \colhead{$\theta_{\rm LD,inf}$} & \colhead{$\sigma_{\rm LD}$} & \colhead{$R$} & \colhead{$L$}   & \colhead{$F_{\rm BOL}$}                      		      & \colhead{$T_{\rm eff}$} & \colhead{$\sigma_{\rm Teff}$} & \colhead{} \\ 
           \colhead{HD}     & \colhead{(mas)}                  & \colhead{Initial}               & \colhead{Final}                & \colhead{(mas)}                 & \colhead{($\%$)}            & \colhead{($R_\odot$)}      & \colhead{($L_{\odot}$)} & \colhead{(10$^{-8}$ erg s$^{-1}$ cm$^{-2}$)} & \colhead{(K)}           & \colhead{$\%$}              & \colhead{$A_{\rm V}$} }
\startdata
2774   & 1.269$\pm$0.023 & 0.32 & 0.31 & 1.303$\pm$0.023 & 1.8 & 16.36$\pm$0.84 & 125.1$\pm$15.6 & 29.3$\pm$2.3 & 4771$\pm$104 & 2 & 0.26$\pm$0.05 \\
6497   & 0.715$\pm$0.044 & 0.33 & 0.29 & 0.731$\pm$0.044 & 6.0 & 7.79$\pm$0.62  & 46.7$\pm$5.4   & 15.2$\pm$0.8 & 5405$\pm$177 & 3 & 0.38$\pm$0.03 \\
13982  & 1.140$\pm$0.032 & 0.33 & 0.31 & 1.169$\pm$0.032 & 2.7 & 15.85$\pm$0.99 & 118.1$\pm$15.1 & 23.8$\pm$1.5 & 4781$\pm$101 & 2 & 0.00$\pm$0.06 \\
31579  & 1.540$\pm$0.008 & 0.35 & 0.36 & 1.593$\pm$0.008 & 0.5 & 30.19$\pm$3.31 & 242.3$\pm$55.2 & 24.9$\pm$1.6 & 4143$\pm$67  & 2 & 0.24$\pm$0.04 \\
153956 & 0.960$\pm$0.023 & 0.33 & 0.29 & 0.983$\pm$0.023 & 2.3 & 9.84$\pm$0.55  & 57.2$\pm$7.8   & 21.1$\pm$1.9 & 5060$\pm$127 & 3 & 0.47$\pm$0.04 \\
157681 & 1.848$\pm$0.013 & 0.32 & 0.34 & 1.908$\pm$0.013 & 0.7 & 39.21$\pm$2.04 & 440.3$\pm$56.7 & 38.5$\pm$3.4 & 4221$\pm$94  & 2 & 0.16$\pm$0.07 \\
184293 & 1.511$\pm$0.022 & 0.33 & 0.29 & 1.548$\pm$0.022 & 1.4 & 23.56$\pm$0.81 & 318.8$\pm$54.4 & 50.8$\pm$8.1 & 5022$\pm$203 & 4 & 0.92$\pm$0.04 \\
216174 & 1.556$\pm$0.012 & 0.34 & 0.32 & 1.598$\pm$0.012 & 0.8 & 20.92$\pm$0.66 & 175.0$\pm$15.5 & 37.7$\pm$2.4 & 4588$\pm$76  & 2 & 0.49$\pm$0.03 \\
218029 & 1.809$\pm$0.044 & 0.34 & 0.33 & 1.862$\pm$0.044 & 2.4 & 25.36$\pm$0.93 & 227.1$\pm$17.8 & 45.2$\pm$2.5 & 4448$\pm$81  & 2 & 0.22$\pm$0.04 \\
\enddata
\tablecomments{$\mu_{\rm \lambda}$ values are from \citet{2011AandA...529A..75C};
$A_{\rm V}$ values are from the SED fits.}
\end{deluxetable}

\clearpage


\begin{deluxetable}{ccccccccc}
\tablewidth{0pc}
\tablecaption{Stellar Parameters Using Various Techniques.\label{tcompare_table}}

\tablehead{\colhead{Target} & \colhead{ }       & \colhead{ }      & \colhead{$T_{(B-V)}$} & \colhead{$T_{\rm spec}$} & \colhead{$T_{\rm inf}$} & \colhead{$T_{\rm lit}$} & \colhead{$T_{(V-K)}$} & \colhead{$\theta_{(V-K)}$} \\
           \colhead{HD}     & \colhead{($B-V$)} & \colhead{BC$_V$} & \colhead{(K)}         & \colhead{(K)}            & \colhead{(K)}           & \colhead{(K)}           & \colhead{(mas)}       & \colhead{(K)}             }
\startdata
2774   & 1.17 & -0.60 & 4300 & 4655 & 4771 & 4524 & 4538 & 1.40$\pm$0.21 \\
6497   & 1.20 & -0.64 & 4245 & 4420 & 5405 & 4433 & 4859 & 0.82$\pm$0.28 \\
13982  & 1.20 & -0.64 & 4010 & 4580 & 4781 & 4678 & 4310 & 1.38$\pm$0.40 \\
31579  & 1.49 & -1.21 & 3900 & 4500 & 4143 & 4154 & 3952 & 1.67$\pm$0.27 \\
153956 & 1.17 & -0.60 & 4300 & 4510 & 5060 & 4636 & 4733 & 1.11$\pm$0.14 \\
157681 & 1.49 & -1.21 & 3900 & 4400 & 4221 & 4164 & 4012 & 2.03$\pm$0.28 \\
184293 & 1.29 & -0.77 & 4100 & 4380 & 5022 & 4465 & 4940 & 1.51$\pm$0.22 \\
216174 & 1.17 & -0.60 & 4300 & 4300 & 4588 & 4488 & 4766 & 1.48$\pm$0.24 \\
218029 & 1.27 & -0.74 & 4150 & 4360 & 4448 & 4362 & 4529 & 1.63$\pm$0.22 \\
\enddata
\tablecomments{The bolometric correction BC$_V$ was calculated using the ($B-V$) color from SIMBAD; $T_{(B-V)}$, $T_{\rm spec}$, $T_{\rm inf}$, $T_{\rm lit}$, and $T_{(V-K)}$ are the temperatures derived from the ($B-V$) color, spectroscopy, interferometry, averaging over temperatures found in the literature using the VizieR service, and the ($V-K$) color, respectively. $\theta_{(V-K)}$ is the angular diameter predicted using the relations in \citet{2003AJ....126.2502M}.}
\end{deluxetable}

\clearpage


\clearpage

\clearpage

\clearpage

\end{document}